\documentclass[twocolumn]{emulateapj}

\usepackage{amsmath,natbib, appendix}
\usepackage{longtable}
\usepackage{graphicx}

\slugcomment{Submitted to ApJ}

\shorttitle{HALO IN DARK AGES}

\shortauthors{NOVOSYADLYJ ET AL.}

\begin{document}
\title{Halos in Dark Ages: formation and chemistry}

\author{B. Novosyadlyj$^{1,2,4}$, V. Shulga$^{1,3,4}$, W. Han$^{1,4}$, Yu. Kulinich$^2$, M. Tsizh$^2$}
 
\affil{$^1$International Center of Future Science of Jilin University, Qianjin Street 2699, Changchun, 130012, P.R.China, \\
$^2$Astronomical Observatory of Ivan Franko National University of Lviv, Kyryla i Methodia str., 8, Lviv, 79005, Ukraine,\\
$^3$Institute of Radio Astronomy of NASU, 4 Mystetstv str., 61002 Kharkiv, Ukraine,\\
$^4$College of Physics of Jilin University, Qianjin Street 2699, Changchun, 130012, P.R.China}

\begin{abstract}
Formation of halos in the Dark Ages from initial spherical perturbations is analyzed in a four component Universe (dark matter, dark energy, baryonic matter and radiation) in the approximation of relativistic hydrodynamics. Evolution of density and velocity perturbations of each component is obtained by integration of a system of nine differential equations from $z=10^8$ up to virialization, which is described phenomenologically. It is shown that the number density of dark matter halos with masses $M\sim10^8-10^9\,\mathrm{M_{\odot}}$ virialized at $z\sim10$ is close to the number density of galaxies in comoving coordinates. The dynamical dark energy of classical scalar field type does not significantly influence the evolution of the other components, but dark energy with a small value of effective sound speed can affect the final halo state. Simultaneously, the formation/dissociation of the first molecules have been analyzed in the halos which are forming. The results show that number densities of molecules $\rm{H_2}$ and $\rm{HD}$ at the moment of halo virialization are $\sim10^3$ and $\sim400$  times larger, respectively, than on a uniformly expanding background. It is caused by increased density and rates of reactions at quasilinear and nonlinear evolution stages of density and velocity of the baryonic component of halos. It is shown also that the temperature history of the halo is important for calculating the concentration of molecular ions with low binding energy. So, in a halo with virial temperature $\sim10^5$ K the number density of the molecular ion HeH$^+$ is approximately 100 times smaller than that on the cosmological background.
 
\end{abstract}

\section{Introduction}

Molecules in the Dark Ages are an important subject of study for a few reasons. First of all, they are effective coolers of collapsing gas in the processes of the first stars formation. 
Without scrupulous account of their role in these processes, we cannot be sure that we know when the first sources of light appeared and which they were.
Secondly, molecules are able to scatter and absorb the quanta of cosmic microwave background, influence their energy and spatial distributions. They can be detected in the next generation CMB experiments. At last the Dark Ages molecules can be a source of light from the Dark Ages which brings the new information about that epoch.

The growing list of molecules and possible reactions can be found in \citet{Dubrovich1977,Izotov1984,Lepp1984,Dalgarno1987,Puy1993,Galli1998,Stancil1998,Puy2002,Hirata2006,Vonlanthen2009,Safranek2010,Gay2011}. The current list of Dark Ages ingredients contains $\sim30$ species and $\sim250$ reactions (see reviews \citet{Lepp1998,Lepp2002,Galli2013}). The main knowledge about primordial chemistry follows from the computations of cosmological recombination of Hydrogen, Deuterium and Helium and formation of neutral and ion molecules after recombination and before re-ionization by the first stars and galaxies ($10\le z\le 1000$) in the $\Lambda$CDM model. It was shown that the number density of molecules in the Dark Ages crucially depends on the number density of free electrons and protons. The evaluation of ionized fractions depends on the accuracy of computation of all atomic and photonic processes during cosmological recombination, existence of additional sources of ionization (e.g. decaying, annihilating dark matter particles etc.) and rate of expansion of the Universe, which depends on the assumption about nature of dark energy. It was shown also that only simple diatomic and triatomic molecules and molecular ions containing H, D, He, and Li are formed in trace amounts during the Dark Ages. Our recent computation for $\Lambda$CDM model with Planck parameters \citep{Planck2015} in the case of absence of sources of re-ionization before $z=10$ shown that the relative number densities (in the units of Hydrogen abundance) are $2.4\times10^{-6}$ for molecule $\rm H_2$, $1.1\times10^{-9}$ for ${\rm HD}$, $1.3\times10^{-13}$ for ${\rm H_2^+}$ and $8.5\times10^{-14}$ for ${\rm HeH^+}$ \citep{Novosyadlyj2017c}. It was found also that the uncertainties of molecular abundances caused by the inaccuracies of computation of cosmological recombination are about 2-3\%. The uncertainties of values of cosmological parameters affect the abundances of molecules at the level of up to 2\%. 

The primordial molecules allow the gas to cool, contract and fragment, that is very important for estimation of mass function of the first luminous objects. Since the cosmological perturbations exist and evolve, the concentrations of the molecules change over time in different places in different ways. They are determined by the dynamics of change of baryonic density, temperature, radiation spectrum and intensity and by the dependence of effective cross sections of molecular formation/destruction reactions on the physical state of the baryonic matter. The number densities of molecules decrease on a uniformly expanding background (further cosmological background or c.b.) and increase in the clouds which contract and virialize. We can compute the evolution of molecular number density on the cosmological background with accuracy which is defined by accuracies of cosmological parameters (a few percents, as it was mentioned above) and accuracies of cross-sections, which are known a bit worse. At the same time the computation of molecular fractions in the forming halos is less certain, since it is dependent on models of halo formation, which are mostly phenomenological in different aspects (see \citet{Lepp1998,Barkana2001,Padmanabhan2002,Lepp2002,Bromm2011,Galli2013} and citing therein).            

In this paper, we study formation of the first molecules in the halos which virialize at the end of the Dark Ages and compare molecular number densities with ones on the cosmological background. We describe the evolution of spherical perturbations in the multicomponent medium from the early stage when the corresponding peaks in the Gaussian field of cosmological perturbations as their seeds were super-horizon up to virialization of dark matter halos. Two models of baryonic gas behavior are considered: (i) when it is adiabatic at all stages and (ii) when it reaches the virial temperature after virialization of dark matter halo. In the second section, we describe the model of spherical scalar perturbations in the four-component medium (dark matter, dark energy, baryonic gas and thermal relic radiation), equations, initial conditions, method of integration and evolution of density and velocity perturbation amplitudes in the central part of spherical overdensities with baryonic mass $\sim10^8-10^9$ $M_{\odot}$. In the third section, we analyze the formation of molecules in the halo and on the cosmological background during the Dark Ages before re-ionization by first stars and compare them. Discussions and conclusions are presented in the fourth section.

\section{Formation of halos in the Dark Ages}

Let us analyze the evolution of spherical perturbations of galaxy scales in the four ingredient Universe (cold dark matter, baryonic matter, dark energy and thermal relic radiation) from a linear stage in the early epoch, through quasi-linear stage, turnaround point, infall and formation of spherical halo at the end of the Dark Ages. We use the hydrodynamical approximation in which each ingredient is a continuous medium with energy density $\varepsilon_{\rm (N)}(t,r)$, pressure $p_{\rm (N)}(t,r)$, equation of state (EoS) parameter $w_{\rm (N)}\equiv p_{\rm (N)}(t,r)/\varepsilon_{\rm (N)}(t,r)$ and squared effective sound speed $c_{s\,(N)}^2\equiv\delta p_{\rm (N)}/\delta\varepsilon_{\rm (N)}$. The cold dark matter, as usual, has dust-like parameters: $w_{\rm dm}=c_{\rm s\,(dm)}^2=0$. For the dark energy, we assume $w_{\rm de}=const=-0.9$ and $c_{\rm s(de)}^2=1$ or $\ll1$ (in units of the speed of light). The baryon-photon plasma at the radiation-dominated epoch is ultra-relativistic with $w_{\rm r,b}=c_{\rm s\,(r,b)}^2=1/3(1+R)$, where $R\equiv3\varepsilon_{\rm b}/4\varepsilon_{\rm r}$ \citep{Hu1995,Hu1996}. After cosmological recombination the baryonic ingredient is practically neutral ideal gas with $w_{\rm b}=c_{\rm s(b)}^2=\tilde{\mathrm{\gamma}} k T_{\rm b}/\mu_{\rm H} m_{\rm H}$, where $T_{\rm b}$ is its temperature, $m_{\rm H}$ is hydrogen atom mass, $\mu_{\rm H}$ is mass per H atom, $k$ is Boltzmann constant and $\tilde{\mathrm{\gamma}}$ is the adiabatic index which we suppose here is equal 5/3. We assume the standard primordial chemistry following from the cosmological nucleosynthesis with helium nucleon fraction $Y_p=0.2465\pm0.0097$ \citep{Aver2013}. In the expanding homogeneous Universe the density of ingredients follow the energy-momentum conservation law
\begin{eqnarray}
&&\bar{\varepsilon}_{\rm dm}(t)=\varepsilon_{\rm dm}^{(0)}a^{-3}, \quad \bar{\varepsilon}_{\rm b}(t)=\varepsilon_{\rm b}^{(0)}a^{-3}, \nonumber\\ 
&&\bar{\varepsilon}_{\rm de}(t)=\varepsilon_{\rm de}^{(0)}a^{-3(1+w_{\rm de})}, \quad \bar{\varepsilon}_{\rm r}(t)=\varepsilon_{\rm r}^{(0)}a^{-4}, \label{eps_0}
\end{eqnarray}
where $a(t)$ is the scale factor normalized to 1 at the current moment of time $t_0$ ($a(t_0)=1$) and $\varepsilon_{\rm (N)}^{(0)}$ is the energy densities at the current epoch which are usually parametrized by the dimensionless density parameters $\Omega_{\rm (N)}=\bar{\varepsilon}_{\rm (N)}^{(0)}/\varepsilon_{\rm (cr)}^{(0)}$, where the critical density 
$\varepsilon_{\rm (cr)}^{(0)}\equiv3c^2H_0^2/8\pi G$. We will use also notation $\Omega_{\rm m}\equiv\Omega_{\rm dm}+\Omega_{\rm b}$. In the computations we assume the Hubble constant $H_0=70$ km/s/Mpc and zero space curvature.

\subsection{Equations of evolution for amplitudes of spherical perturbation and initial conditions}

We suppose that halo forms from spherical adiabatic perturbation in which density of each ingredient is  $\varepsilon_{\rm (N)}(t,r)=\bar{\varepsilon}_{\rm (N)}(t)(1+\delta_{\rm (N)}(t,r))$ and Friedmann-Lemaitre-Robertson-Walker metric is slightly perturbed  
\begin{equation}
ds^2=e^{\nu(t,r)}dt^2-a^2(t)e^{\nu(t,r)}[dr^2+r^2d\Omega^2],  \label{ds} 
\end{equation}
where $d\Omega^2=d\theta^2+\sin^2\theta d\varphi^2$. In the early epoch $\delta_{\rm (N)}\ll1$ and $\nu\ll1$. The components of four-velocity $u^{\rm i}_{\rm (N)}(t,r)\equiv dx^{\rm i}_{\rm (N)}/ds$ is connected with components of three-velocity $v_{\rm (N)}^{\rm i}(t,r)\equiv dr_{\rm (N)}/d\tau$, where $d\tau\equiv e^{\nu(t,r)}dt$ is an interval of proper time of observer in the point of space-time $(t,r)$, by relations    
\begin{equation}
u^{\rm i}_{\rm (N)}(t,r) = \left\{\frac{e^{-\nu/2}}{\sqrt{1-v_{\rm (N)}^2}},\,\,\frac{e^{-\nu/2}v_{\rm (N)}}{a\sqrt{1-v_{\rm (N)}^2}},\,\,0,\,\,0\right\}.\nonumber
\end{equation}
The components of energy-momentum tensor of ideal fluid in $v^2$-approximation are as follows: 
\begin{eqnarray} 
&&T^0_{\rm 0\,(N)} = \varepsilon_{\rm (N)}+(\varepsilon_{\rm (N)}+p_{\rm (N)})v_{\rm (N)}^2, \nonumber\\
&&T^1_{\rm 0\,(N)} = a^{-1}(\varepsilon_{\rm (N)}+p_{\rm (N)})v_{\rm (N)}, \nonumber\\
&&T^1_{\rm 1\,(N)} = -p_{\rm (N)}-(\varepsilon_{\rm (N)}+p_{\rm (N)})v_{\rm (N)}^2 ,\nonumber\\  
&&T^2_{\rm 2\,(N)} = T^3_{\rm 3\,(N)} = - p_{\rm (N)}. \nonumber  
\end{eqnarray} 
The evolution of the cosmological background and spherical perturbation is described by Einstein equations of general relativity and equations of energy-momentum conservation 
\begin{eqnarray}
R^{\rm i}_{\rm j}-\frac{1}{2}\delta^{\rm i}_{\rm j} R=\frac{8\pi G}{c^4} \sum_{\rm N} T^{\rm i\;(N)}_{\rm j}, \quad  T^{\rm k\;(N)}_{\rm i\,;k}=0. \label{main_eq}
\end{eqnarray}
For the cosmological background ($\nu=\delta_{\rm (N)}=v_{\rm (N)}=0$) they give the Friedmann equations
\begin{eqnarray}
H&=&H_0\sqrt{\Omega_{\rm r} a^{-4}+\Omega_{\rm m} a^{-3}+\Omega_{\rm de}a^{-3(1+w_{\rm de})}}, \\ 
q&=&\frac{\Omega_{\rm r} a^{-4}+\Omega_{\rm m} a^{-3}/2+(1+3w_{\rm de})\Omega_{\rm de}a^{-3(1+w_{\rm de})}}{\Omega_{\rm r} a^{-4}+\Omega_{\rm m} a^{-3}+\Omega_{\rm de}a^{-3(1+w_{\rm de})}},\nonumber
\end{eqnarray} 
which describe the dynamic of expansion of the Universe in the terms of its rate $H\equiv d\ln{a}/dt$ and deceleration parameters $q\equiv-d^2a/dt^2/H^2$. The integration of the first one from them gives the dependence $a(t)$ or/and $t(a)$. It is comfortable to use $a$ as independent variable instead $t$, since it is simply connected with redshift
$z=1/a-1$, which is measurable value. The derivative with respect to $t$ is presented over the derivative with respect to $a$ as follows $\frac{d}{dt}=aH\frac{d}{da}$. For compactness below, we will use the notation: $\left(\dot{}\right)\equiv\frac{d}{da}$. 

The perturbed part of equations \ref{main_eq} consists from the system of nine equations for nine unknown functions of $a$ and $r$: $\nu(a,r)$, $\delta_{\rm dm}(a,r)$, $v_{\rm dm}(a,r)$, $\delta_{\rm b}(a,r)$, $v_{\rm b}(a,r)$, $\delta_{\rm de}(a,r)$, $v_{\rm de}(a,r)$, $\delta_{\rm r}(a,r)$, $v_{\rm r}(a,r)$. In this paper we are interested in mainly the densest central part of a halo formed in the Dark Ages. At the early epoch, the perturbations are small and equations of their evolution can be linearized for all ingredients. Moreover, each function of $a$ and $r$ can be presented as a product of its amplitude, which depends on $a$ only, and some function of radial coordinate $r$, which describes the initial profile of spherical perturbation. The last can be expanded into series of some orthogonal functions, e.g. spherical ones in our case. In particular, we can present the perturbations of the metric, density and velocity of N-ingredient as follows
\begin{eqnarray}
&&\nu(a,r)=\tilde{\nu}(a)\frac{\sin{kr}}{kr}, \quad \delta_{\rm (N)}(a,r)=\tilde{\delta}_{\rm (N)}(a)\frac{\sin{kr}}{kr}, \nonumber \\
&&v_{\rm (N)}(a,r)=\tilde{v}_{\rm (N)}(a)k\left(\frac{\cos{kr}}{kr}-\frac{\sin{kr}}{k^2r^2}\right). \label{functions}
\end{eqnarray}
For analyzing the evolution of central part of spherical halo, we can decompose $r$-function in the Taylor series and keep only leading terms:
$$f_k(r)\approx 1,\quad f_k'(r)\approx -\frac{1}{3}k^2 r, \quad f_k''(r) + \frac{2f_k(r)}{r} \approx -k^2,$$
where $f_k(r)=\sin{kr}/kr$. It gives the possibility to reduce the system of nine partial differential equation for unknown functions
(\ref{functions}) to the system of nine ordinary differential equations for their amplitudes $\tilde{\nu}(a)$, $\tilde{\delta}_{\rm dm}(a)$, $\tilde{v}_{\rm dm}(a)$, 
$\tilde{\delta}_{\rm b}(a)$, $\tilde{v}_{\rm b}(a)$, $\tilde{\delta}_{\rm de}(a)$, $\tilde{v}_{\rm de}(a)$, $\tilde{\delta}_{\rm r}(a)$, $\tilde{v}_{\rm r}(a)$:
\begin{widetext}
\begin{eqnarray}
&&\hskip-1.5cm\dot{\tilde{\nu}}+\left(1+(1-\tilde{\nu})\frac{k^2}{3a^2H^2}\right)\frac{\tilde{\nu}}{a} = 
-\frac{\Omega_{\rm m}\tilde{\delta}_{\rm m}+\Omega_{\rm r}a^{-1}\tilde{\delta}_{\rm r}+\Omega_{\rm de}a^{-3w_{\rm (N)}}\tilde{\delta}_{\rm de}}
{\Omega_{\rm m}a+\Omega_{\rm r}+\Omega_{\rm de}a^{1-3w_{\rm (N)}}}, \label{nu} \\
&&\hskip-1.5cm\dot{\tilde{\delta}}_{\rm (N)}+\frac{3}{a}(c^2_{s(N)}-w_{\rm (N)})\tilde{\delta}_{\rm (N)}-(1+w_{\rm (N)})\left[\frac{k^2\tilde{v}_{\rm (N)}}{a^2H}
+9H(c^2_{s(N)}-w_{\rm (N)})\tilde{v}_{\rm (N)}+\frac{3}{2}\dot{\tilde{\nu}}\right]- \nonumber \\
&&\hskip8.5cm-(1+c^2_{s(N)})\left[\frac{k^2\tilde{\delta}_{\rm (N)}\tilde{v}_{\rm (N)}}{a^2H}
+\frac{3}{2}\tilde{\delta}_{\rm (N)}\dot{\tilde{\nu}}\right]=0, \label{delta} \\
&&\hskip-1.5cm\dot{\tilde{v}}_{\rm (N)}+(1-3c^2_{s(N)})\frac{\tilde{v}_{\rm (N)}}{a}+\frac{c^2_{s(N)}\tilde{\delta}_{\rm (N)}}{a^2H(1+w_{\rm (N)})}+
\frac{\tilde{\nu}}{2a^2H}-\frac{4k^2\tilde{v}^2_{\rm (N)}}{3a^2H}+\nonumber\\
&&\hskip4cm +\frac{1+c^2_{s(N)}}{1+w_{\rm (N)}}\left[\dot{\tilde{\delta}}_{\rm (N)}\tilde{v}_{\rm (N)}+\tilde{\delta}_{\rm (N)}\dot{\tilde{v}}_{\rm (N)}+
(1-3w_{\rm (N)})\frac{\tilde{\delta}_{\rm (N)}}{a}\tilde{v}_{\rm (N)}+\frac{\tilde{\nu}\tilde{\delta}_{\rm (N)}}{2a^2H}\right]=0. \label{v}
\end{eqnarray}
\end{widetext}
To take into account the Silk damping and drag effects for the baryon-photon plasma before recombination we supplement the right hand sides of eqs. (\ref{delta})-(\ref{v}) for photons by terms 
$-k^2\tilde{\delta}_{\rm r}/a^2Hk_{\rm D}^2$ and $-k^2\tilde{v}_{\rm r}/a^2Hk_{\rm D}^2$ and for baryons $-k^2\tilde{\delta}_{\rm b}/a^2Hk_{\rm D}^2e^{k^2/k_{\rm D}^2}$ and 
$-k^2\tilde{v}_{\rm b}/a^2Hk_{\rm D}^2e^{k^2/k_{\rm D}^2}$ correspondingly. The damping scale $k_{\rm D}$ we compute semi-analytically according to \citet{Hu1995}.

To integrate them we should set the initial conditions for each function at radiation-dominated epoch at $a_{\rm init}=10^{-8}$ when cluster and galaxy scale perturbations are super-horizon. Linearized equations (\ref{nu})-(\ref{v}) for radiation component has asymptotic values at $a_{\rm init}$
\begin{equation}
\tilde{\nu}^{\rm init}=-C_k, \quad \tilde{\delta}^{\rm init}_{\rm r}=C_k, \quad \tilde{v}^{\rm init}_{\rm r}=\frac{C_k}{4a_{\rm init}H(a_{\rm init})}, \label{r_ini}  
\end{equation}
where $C_k$ is some constant. The solutions of equations (\ref{nu})-(\ref{v}) for matter and dark energy as the test components give the asymptotic values for super-horizon perturbations at 
$a_{\rm init}$
\begin{eqnarray}
&&\tilde{\delta}_{\rm dm}^{\rm init}=\tilde{\delta}_{\rm b}^{\rm init}=\frac{3}{4}C_k,  \quad  \tilde{\delta}_{\rm de}^{\rm init}=\frac{3}{4}(1+w_{de})C_k, \nonumber \\ 
&&\tilde{v}_{\rm m}^{\rm init}=\tilde{v}_{\rm b}^{\rm init}=\tilde{v}_{\rm de}^{\rm init}=\frac{C_k}{4a_{\rm init}H(a_{\rm init})}. \label{m_ini}
\end{eqnarray}
We will see that $k\sim1-10$ Mpc$^{-1}$ scale perturbations can form the halo at $z\sim30-10$ if their amplitudes $C_k\sim (1-3)\cdot10^{-4}$. Let us compare these values  with the root mean square (rms) amplitude followed from the initial power spectrum normalized to Planck2015 data \citep{Planck2015}. 
\begin{figure*}
\includegraphics[width=0.49\textwidth]{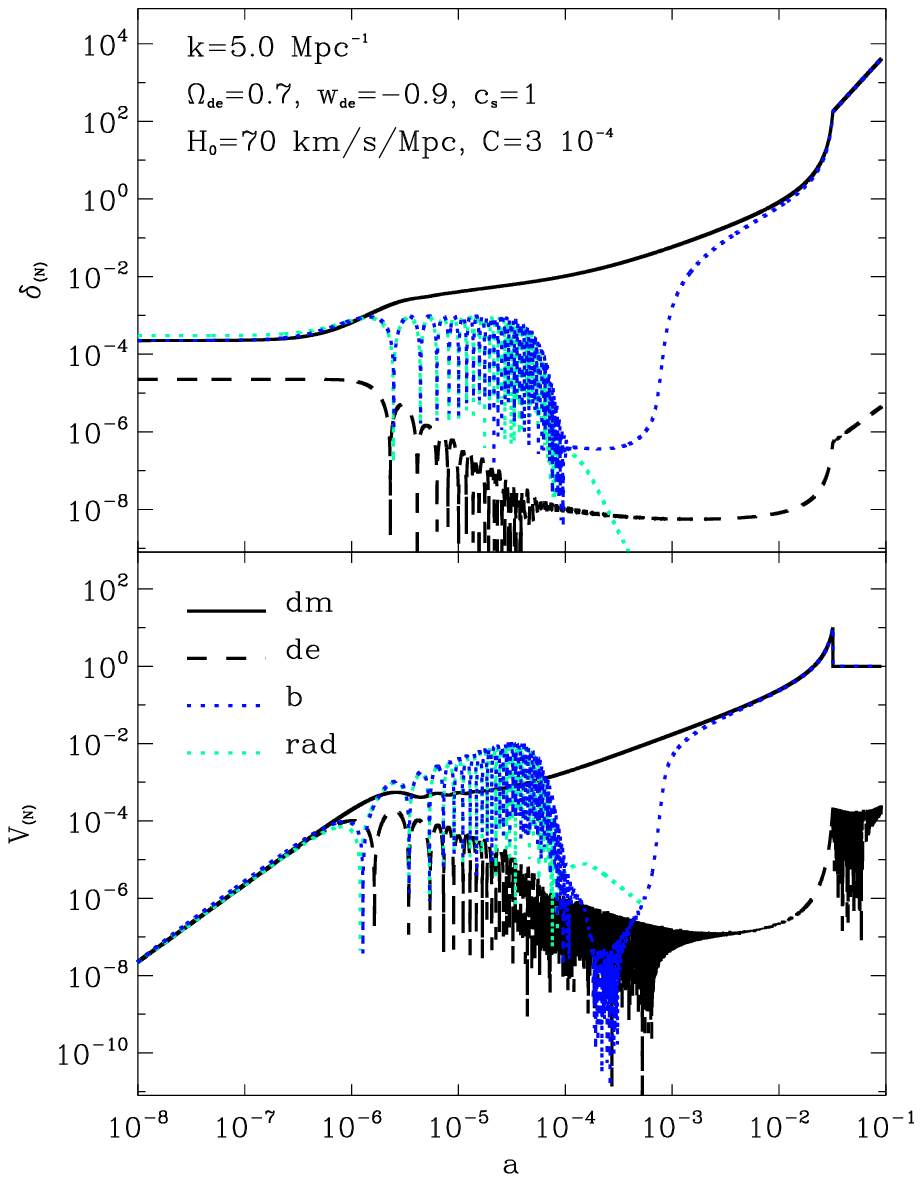}
\includegraphics[width=0.49\textwidth]{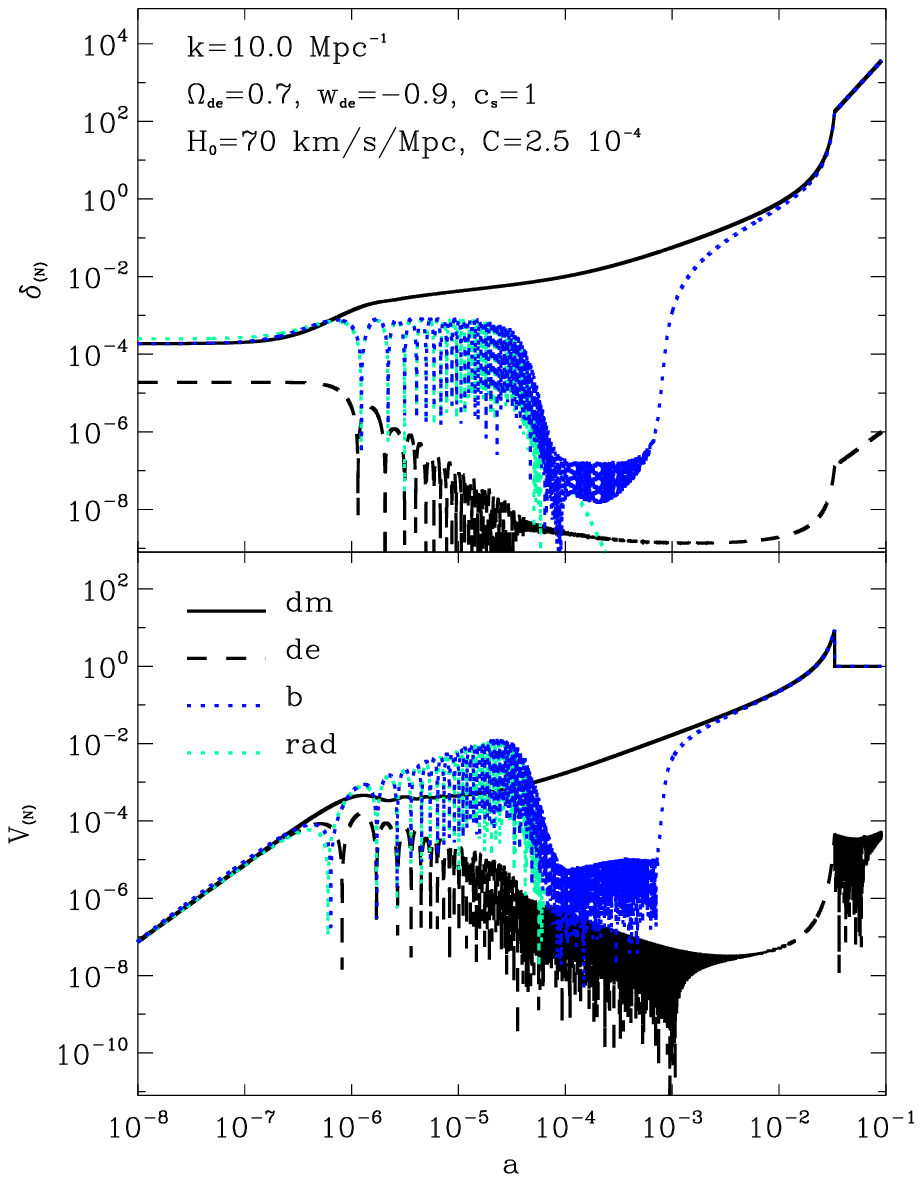}
\caption{Evolution of density and velocity perturbations of the matter, dark energy, baryons and radiation. Cosmological parameters here and below are as follows: $H_0=70$ km/s/Mpc, $\Omega_{\rm de}=0.7$, $w_0=-0.9$, $c_{\rm s(de)}^2=1$, $\Omega_{\rm b}=0.05$, $\Omega_K=0$, $T_0=2.7255$ K. }
\label{vir}
\end{figure*}

The power spectrum of curvature perturbations usually is presented as $\mathcal{P}_{R}(k)=A_{\rm s}(k/0.05)^{n_{\rm s}-1}$ \citep{camb}, where $n_{\rm s}$ is spectral index of scalar mode of cosmological density perturbations and $A_{\rm s}$ is its amplitude at $k=0.05$ Mpc$^{-1}$. Using the Planck2015 + HST + WiggleZ + SNLS3 data set we have determined their mean values and 2$\sigma$ confidence limits: $A_{\rm s}=2.19^{+0.12}_{-0.10}\cdot10^{-9},\,\,n_{\rm s}=0.960\pm0.013$ \citep{Sergijenko2015}. For the perturbations of super-horizon scales $\mathcal{P}_{R}\equiv\frac{9}{16}<\tilde{\nu}\cdot\tilde{\nu}>=const$ the rms value of metric perturbation $\sigma_{\nu}\equiv<\tilde{\nu}\cdot\tilde{\nu}>^{1/2}\approx5.7\cdot10^{-5}\left(k/5\right)^{\frac{n_{\rm s}-1}{2}}.$ It means that the height of peaks in the Gaussian random field of matter density perturbations is ~2-5 $\sigma_{\nu}$. 

The system of equations (\ref{nu})-(\ref{v}) should be supplemented by equations of evolution of temperature of the gas, since it determines the pressure and effective sound speed. At $z>850$ the temperature of baryonic matter $T_{\rm b}$ practically equals to the temperature of radiation $T_{\rm r}$ which adiabatically cools with expanding of the Universe:
\begin{equation}
  \label{Tm1}
  \frac{d\;T_{\rm b}}{d\;z}=\frac{T_{\rm b}}{1+z}.
\end{equation}
So, at $a_{\rm init}\le a\le 0.0012$ ($850< z\le z_{\rm init}$) $T_{\rm b}\approx T_{\rm r}=T_0a^{-1}$, where $T_0=2.7255\pm0.0006$ K \citep{Fixsen2009}.
At the lower redshifts, when $z\lesssim850$, the baryonic matter is slightly connected with thermal background radiation via its Compton scattering on the residual fraction of free electrons. Since the cooling rate function of baryonic gas with primordial chemistry is negligible (see Fig. 1 in \citet{Safranek2010}), the adiabatic cooling/heating of mono-molecular gas, at last, becomes the main thermal process for it. At this stage the temperature of the baryonic matter can be described by the next equation:
\begin{eqnarray}
  \label{Tm2}  
\frac{d\;T_{\rm b}}{d\;z}&=&\frac{2T_{\rm b}}{1+z}\left(1+\frac{2}{3}\frac{d\tilde{\delta}_{\rm b}}{dz}\right)   \\ 
&+&\frac{8\sigma_{\rm T}a_{\rm r}T_{\rm r}^4}{3m_{\rm e}cH(1+z)}\frac{
x_{\rm e}}{1+f_{\rm D}+f_{\rm He}+x_{\rm e}}\left(T_{\rm b}-T_{\rm r}\right),\nonumber
\end{eqnarray}
where $x_{\rm e}$, $f_{\rm He}$ and $f_{\rm H}$ are free electrons, total Helium and Deuterium fractions accordingly (see section III for details). The constants are the speed of light $c$, Thomson scattering cross section $\sigma_{\rm T}$, the electron mass $m_{\rm e}$, the radiation constant $a_{\rm r}$.

The system of equations (\ref{nu})-(\ref{v}) and (\ref{Tm1})-(\ref{Tm2}) describes the evolution of density and velocity perturbations of each ingredient with initial conditions (\ref{r_ini})-(\ref{m_ini}) at linear and nonlinear stages, turn around and infall, but without of the final stage of halo formation - virialization. 

\subsection{Formation of halo}

\begin{figure*}
\includegraphics[width=0.49\textwidth]{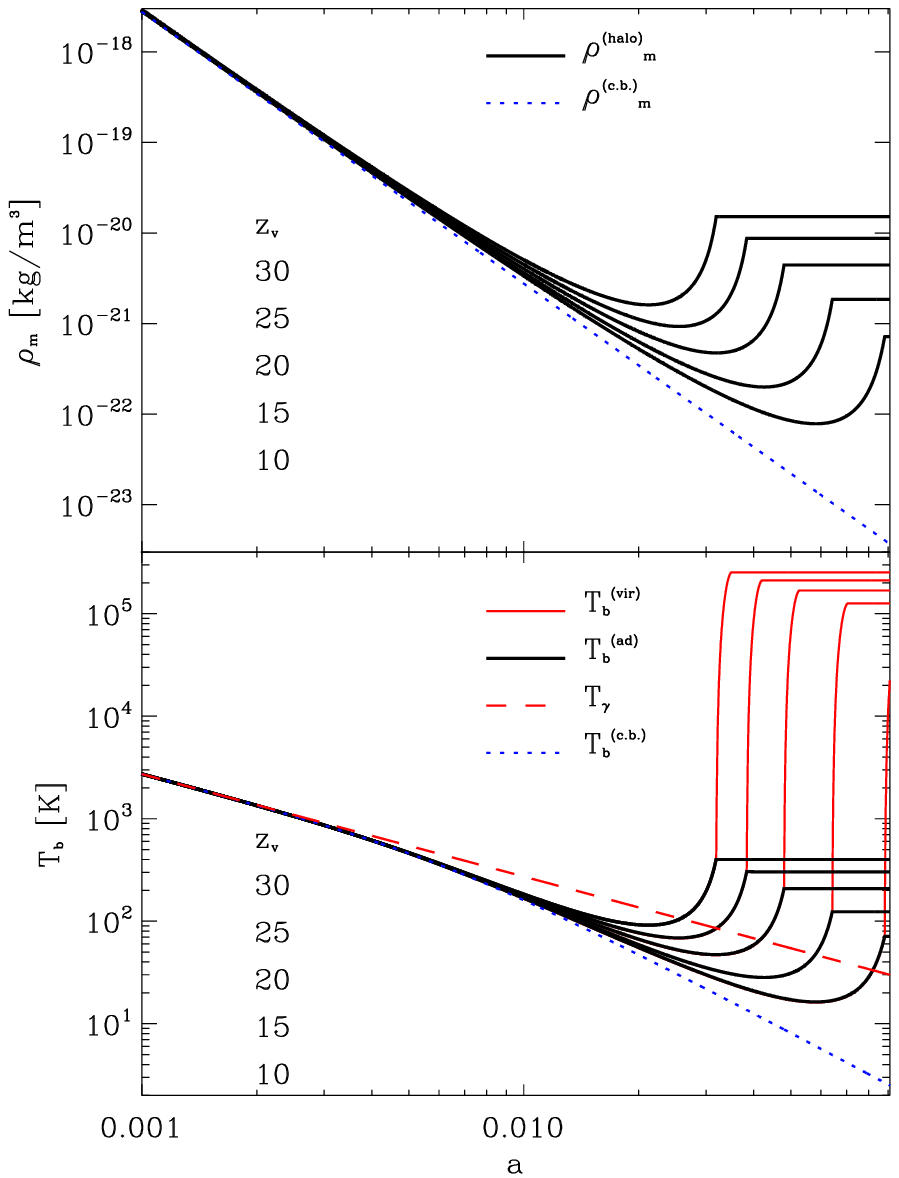} 
\includegraphics[width=0.49\textwidth]{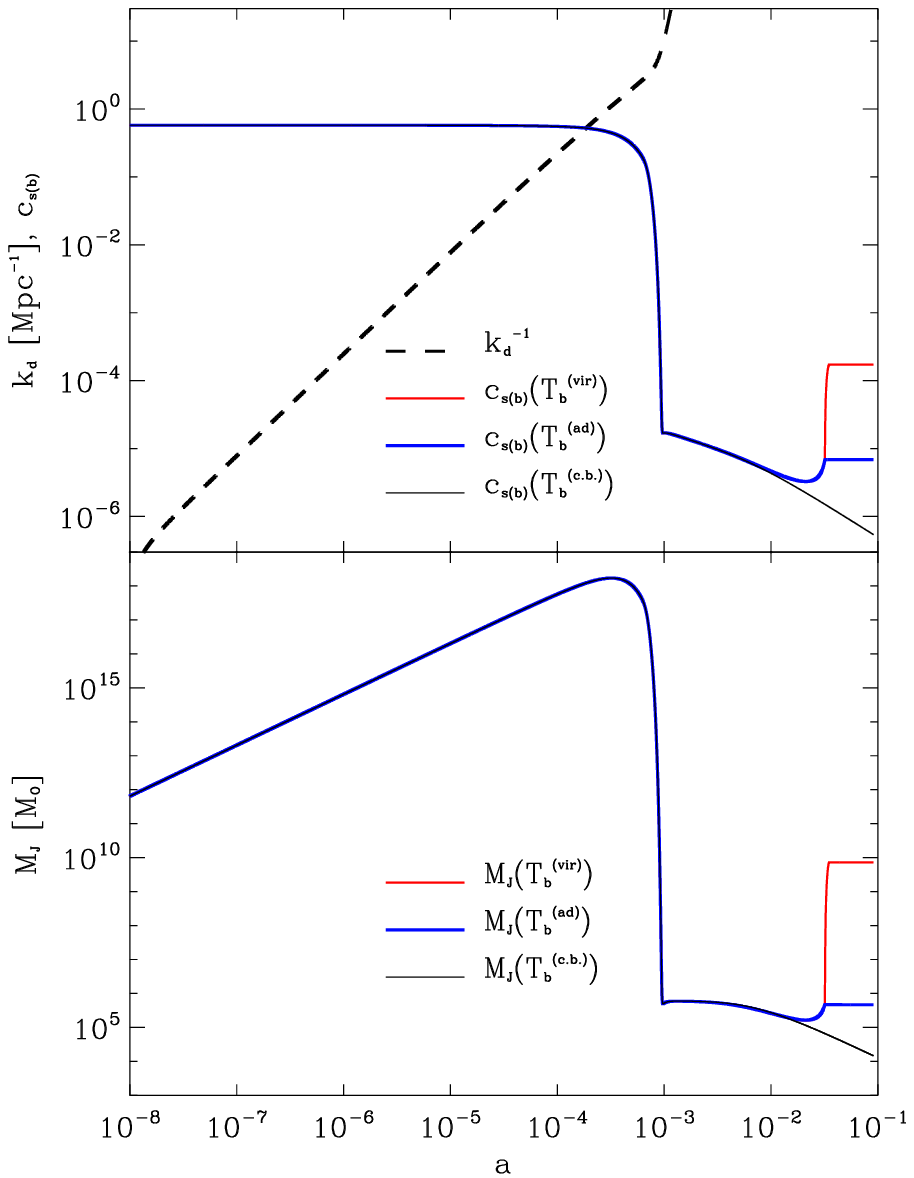}
\caption{Left: Evolution of density of matter $\rho_{\rm m}$ (top panel) and the temperature of baryonic matter $T_{\rm b}$ (bottom panel) for halos with 
$M_{\rm halo}=7\cdot10^8\,\mathrm{M_{\odot}}$ virialized at different redshifts $z_{\rm v}$. The corresponding values on the cosmological background (c.b.) are shown by dotted blue lines. The temperature of thermal relic radiation is shown by dashed red line. Adiabatic temperature of gas is shown by solid dark lines, the virial temperature of gas is shown by red lines. Right: Silk damping and effective sound speed in the baryonic matter before, during and after cosmological recombination (top panel) and Jeans scale for baryonic matter without molecular cooling (bottom panel). The thin solid dark lines show the same for the cosmological background.}
\label{rho-t-sj}
\end{figure*}

Here we will not detail the final processes of the  halo formation such as molecular cooling, fragmentation, star formation and reionization (see, for example, \citep{Oh2002,Safranek2010,Demianski2014}), but will use an approximate description of achievement of a state of dynamic equilibrium based on the virial  theorem. It was shown that in the flat model with $\Omega_{\rm de}\sim0.7$ the overdensity of halo of dark matter $\Delta_{\rm v}\equiv\varepsilon_{\rm dm}(a_{\rm v},r=0)/\bar{\varepsilon}_{\rm dm}(a_{\rm v})-1\approx177$ at the moment of virialization $a_{\rm v}<0.5$  \citep{Kulinich2003}. So, when $\tilde{\delta}_{\rm dm}$ reach $\Delta_{\rm v}$ we assume 
$\tilde{\delta}_{\rm dm}(a\ge a_{\rm v},r=0)=(\Delta_{\rm v}+1)(a/a_{\rm v})^3-1$ and $\tilde{v}_{\rm dm}(a\ge a_{\rm v})=-\tilde{V}_{\rm H}$, where $\tilde{V}_{\rm H}=aH$ is the amplitude of Hubble velocity $V_{\rm H}=aHr$ at the infinitesimal distance $r$ from the center halo (see for details \citep{Novosyadlyj2016}). The baryonic matter during halo formation is first heated adiabatically, and later by shock waves, if they are generated.  

In Fig. \ref{vir} we show the results of integration of (\ref{nu})-(\ref{v}) and (\ref{Tm1})-(\ref{Tm2}) with initial conditions (\ref{r_ini})-(\ref{m_ini}). The dynamical dark energy there is a classical scalar field. The evolution of amplitudes of density perturbation $\tilde{\delta}_{\rm (N)}$ and velocity $\tilde{v}_{\rm (N)}$ of each ingredient up to formation of dark matter halo is presented there. In the left panel the scale of initial perturbation is $\lambda/2=\pi k^{-1}\approx0.6$ Mpc, in the right one $\approx0.3$ Mpc. They form the halos with mass of baryonic matter $\sim10^9$  and $\sim10^8$ M$_\odot$ accordingly at the redshift $\sim30$. The initial amplitudes of metric perturbations for them are $-3\cdot10^{-4}$ and $-2.5\cdot10^{-4}$, which exceed the rms value in $\approx$5.3 and $\approx$4.4 times respectively. Such peaks in the Gaussian random field of curvature perturbations are very rare. If the amplitudes  are $\sim2-3$ times lower then the perturbations of such scales form the halos at redshift $\sim15-10$. The number density of such halos is comparable with the mean number density of galaxies. 

The evolution of mass matter density $\rho_{\rm m}(a)=\rho_{\rm m}^{0}(1+\tilde{\delta}_{\rm m}(a))a^{-3}$ and temperature of baryonic ingredient $T_{\rm b}$ in the central part of halo are shown in Fig. \ref{rho-t-sj} (left column). The temperature of gas in the virialized halo is shown for two important cases: when it is heated by adiabatically compression only $T^{\rm (ad)}_{\rm b}$ (dark thick solid lines) and it is heated by shocks in the processes of violent relaxation to virial temperature $T^{\rm (vir)}_{\rm b}$ (red thick solid lines). In the first case, the final temperature is result of integration of eq. (\ref{Tm2}), in the second one, we set it by hand at $a\ge a_{\rm v}$ from \citet{Barkana2001,Bromm2011}:
\begin{equation}
 T^{\rm (vir)}_{\rm b}=2\cdot10^4\left(\frac{\mu_{\rm H}}{1.2}\right)\left(\frac{M_{\rm halo}}{10^8\mathrm{M_{\odot}}}\right)^{2/3}\left(\frac{\Delta_{\rm v}}{178}\right)^{1/3}\left(\frac{0.1}{a_{\rm v}}\right)\,\mathrm{K}.
\label{Tvir}     
\end{equation} 
We suppose that the temperature of the gas in the real halos of different masses are in the range: $T^{\rm (ad)}_{\rm b}\le T_{\rm b}\le T^{\rm (vir)}_{\rm b}$. One can see, that density and temperature of gas are higher in halos which are formed earlier. This is in agreement with N-body simulations of large-scale structure formation \citep{Klypin2011}.

In the top panel of Fig. \ref{rho-t-sj} (right column) we show how the Silk damping scale $k_{\rm D}$ and the effective sound speed in the baryonic matter (in units of the speed of light $c$) vary with time through epochs. The transition of effective sound speed in the baryonic matter from complete ionized ultrarelativistic stage to practically neutral one in the Dark Ages we described phenomenologically as follows
$$c_{\rm s(b)}^2=\frac{1}{3(1+R)}\left(\frac{n_{\rm HII}}{n_{\rm H}}\right)^2+\frac{\tilde{\mathrm{\gamma}}k T_{\rm b}}{\mu_{\rm H} m_{\rm H} c}\left(\frac{n_{\rm HI}}{n_{\rm H}}\right)^2,$$
where the ionized and neutral fractions of Hydrogen are accurately computed (see next section). 

\begin{figure*}[ht!]
\includegraphics[width=0.49\textwidth]{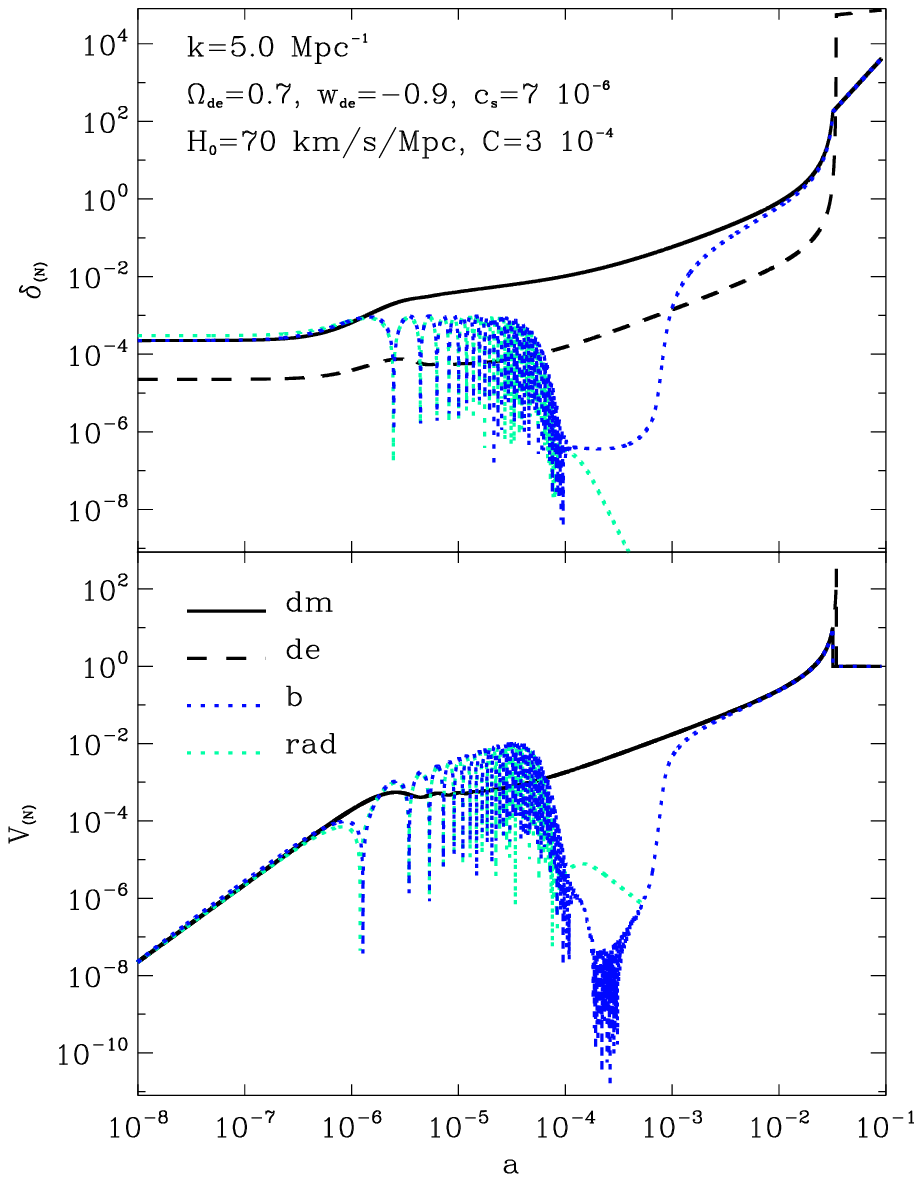}
\includegraphics[width=0.49\textwidth]{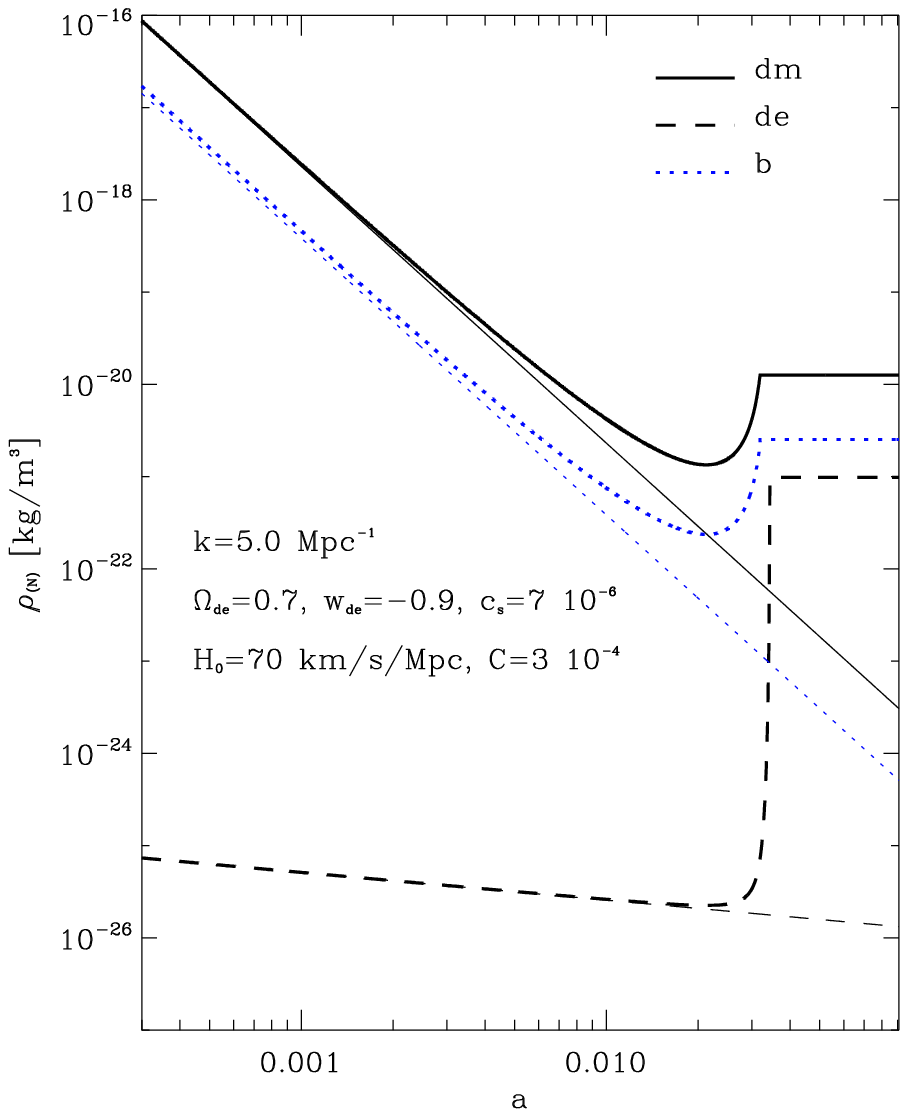}
\caption{Left: Evolution of density and velocity perturbations of matter, dark energy, baryons and radiation which form a halo with virialized matter and dark energy, which reached the equilibrium state when the gravitational force is balanced by pressure gradient. Right: The density of dark matter (solid lines), dark energy (dashed lines) and baryonic gas (blue dotted lines). Thick lines - halo, thin lines - cosmological background.}
\label{vir_st}
\end{figure*}

Now we can estimate the variation of Jeans mass of baryonic matter through epochs on the cosmological background and in the halo. Since the Jeans scale 
$\lambda_{\rm J}=\left(\pi c_{\rm s(b)}^2/G\rho_{\rm b}\right)^{1/2}$ the mass of baryonic matter in the sphere with radius $\lambda_{\rm J}/2$ is as follows
$$M_{\rm J}=7.2\cdot10^{10}\frac{c_{\rm s(b)}^3}{\sqrt\rho_{\rm b}}\,\rm{M}_{\odot},$$
where effective sound speed is in the units of the speed of light and baryonic density is in units ${\rm kg}/ {\rm m}^3$. For $z<800$ this formula gives practically the same values as formula (40) from \citet{Galli2013}. The variation of Jeans mass through epochs is shown in bottom panel of 
Fig. \ref{rho-t-sj} (right column). One can see, that the baryonic matter can not fragment into clumps of stellar mass without additional cooling, the Jeans mass is $>10^4$ M$_\odot$. 

Fig. \ref{vir} has shown that the dark energy of type of classical scalar field is slightly perturbed and practically does not influence the formation of dark matter halo. But the dark energy of other type can be disturbed essentially and can affect on the virialization and final parameters of halo 
\citep{Mota2004,Maor2005,Manera2006,Nunes2006,Creminelli2010,wang2012,Novosyadlyj2016}. Let us show that for the type of dark energy with a low value of effective sound speed\footnote{Cosmological observations practically do not constrain it (see, for example, \citep{Sergijenko2015}).} the result can be other. In the paper \citet{Novosyadlyj2014b} it was shown that the static solution for some type of dark energy exists in the gravitational field of a spherical static halo. Since the ratio ``gravitational radius to radius of halo'' is very small for the cases we are interested here, formula (17) from \citet{Novosyadlyj2014b} for the dark energy density at the center of halo has simple asymptotic for  $c_{\rm s(de)}\ne0$:
\begin{equation}
\frac{ \rho^{\rm st}_{\rm de}}{ \tilde{\rho}_{\rm de}(z_{\rm v})}=1+1.47\cdot10^{-4}\frac{1+w_{\rm de}}{c_{\rm s(de)}^2}(1+z_{\rm v})\Omega_{\rm m}h^2\left({k\,{\rm Mpc}}\right)^{-2}.
 \label{rho_st}
\end{equation}
We show in Fig. \ref{vir_st} the evolution of density $\tilde{\delta}_{\rm de}$ and velocity $\tilde{v}_{\rm de}$ perturbations (left) and density (right) of such type of dark energy with $c_{\rm s(de)}\approx7\cdot10^{-6}$ (that is close to $c_{\rm s(b)}$ at the Dark Ages) in the model with the same cosmological and halo parameters as in the left Fig. \ref{vir}. It is obtained in the following way: the equations (\ref{delta})-(\ref{v}) for dark energy is integrated up to $\rho_{\rm de}$ reaches the static value (\ref{rho_st}), then we put $\rho_{\rm de}= \rho^{\rm st}_{\rm de}$ and $\tilde{v}_{\rm de}=-\tilde{V}_{\rm H}$. It means that after formation of dark matter halo the dark energy continues to inflow until it reaches the equilibrium state when the gravitational force is balanced by pressure gradient. This dark matter halo is formed at $z\approx30.4$, the density of dark energy reaches the static value (\ref{rho_st}) at $z\approx28.1$, so, in $\sim12$ Myr. One can see that for a bit lower values of $c_{\rm s(de)}$ the density of dark energy in halo will be comparable with main Dark Ages ingredients - dark matter and baryons.   

\begin{table*}[!ht]
\caption{Chemical reactions (their notation in brackets corresponds to \citet{Galli1998}) }
\label{tab1}
\centering
\begin{tabular}{l l l l}
\hline
(H1)&$\mathrm{H^+ + e^- \rightarrow H + \gamma}$&(H2)&$\mathrm{H + \gamma 
\rightarrow 
H^+ + e^-}$\\
(H3)&$\mathrm{H + e^- \rightarrow H^- + \gamma}$&(H4)&$\mathrm{H^- + \gamma 
\rightarrow H + e^-}$\\
(H5)&$\mathrm{H^- + H \rightarrow H_2 + e^-}$&(H7)&$\mathrm{H^- + H^+ 
\rightarrow H 
+H}$ \\
(H8)&$\mathrm{H + H^+ \rightarrow H_2^+ + \gamma}$&(H9)&$\mathrm{H_2^+ + \gamma 
\rightarrow H + H^+}$ \\
(H10)&$\mathrm{H_2^+ + H \rightarrow H_2 + H^+}$&(H15)&$\mathrm{H_2 + H^+ 
\rightarrow 
H_2^+ + H}$\\
\hline
(D1)&$\mathrm{D^+ + e^- \rightarrow D + \gamma}$&(D2)&$\mathrm{D + \gamma 
\rightarrow D^+ + e^-}$\\
(D3)&$\mathrm{D + H^+ \rightarrow D^+ + H}$&(D4)&$\mathrm{D^+ + H \rightarrow D+H^+}$\\
(D8)&$\mathrm{D^+ + H_2 \rightarrow H^+ + HD}$&(D10)&$\mathrm{HD + H^+ 
\rightarrow 
H_2 + D^+}$\\
\hline
(He8)&$\mathrm{He + H^+ \rightarrow HeH^+ + \gamma}$&&\\
(He11)&$\mathrm{HeH^+ + H \rightarrow He + H_2^+}$&(He14)&$\mathrm{HeH^+ + 
\gamma 
\rightarrow He + H^+}$\\
\hline
\end{tabular}
\end{table*}

\section{Molecules in the Dark Age halos}

Let us estimate now the number densities of molecules in such halos and compare them with the corresponding values on the cosmological background. For that, we compute the fractions $x_{\rm i}$ of neutral atoms, molecules and ions in the manner described in \citet{Galli1998,Galli2013,Novosyadlyj2017c}. The index ``${\rm i}$'' notes here and below any atom, molecule or their ions presented in Table \ref{tab1}, where are all reactions used here to estimate the chemistry composition of first halos. It is the minimal model which consists of 19 reactions \citep{Galli1998}: ten for Hydrogen, six for Deuterium and three for Helium.

The general form of equation of chemical kinetics is as follows \citep{Puy1993,Galli1998,Vonlanthen2009}:
\begin{eqnarray}
aH\dot{x}_{\rm i}&=&\sum_{\rm mn}k_{\rm mn}^{\rm (i)}f_{\rm m} f_{\rm n} x_{\rm m}x_{\rm n}+\sum_{\rm m}k_{\rm m\gamma}^{\rm (i)}f_{\rm m} x_{\rm m}\nonumber\\
&-&\sum_{\rm j}k_{\rm ij}f_{\rm i} f_{\rm j} x_{\rm i}x_{\rm j}-k_{\rm i\gamma}f_{\rm i}x_{\rm i},
\label{cin}
\end{eqnarray}
 where $k_{\rm mn}^{\rm (i)}$ is reaction rates for the reactants ${\rm m}$ and ${\rm n}$ which lead to formation of atom/molecule/ion ${\rm i}$, $f_{\rm m}$ is $f_{\rm He}=n_{\rm He}/n_{\rm H}$ for reactant ${\rm m}$ containing Helium, $f_{\rm D}=n_{\rm D}/n_{\rm H}$ for reactant ${\rm m}$ containing Deuterium and $f_{\rm H}\equiv1$ for reactant ${\rm m}$ containing Hydrogen only. For chemical species containing only Hydrogen the fraction ${\rm m}$ is $x_{\rm m}=n_{\rm m}/n_{\rm H}$, where $n_{\rm m}$ is the number density of species ${\rm m}$, $n_{\rm H}$ is the total number density of Hydrogen; for species containing Deuterium and Helium  $x_{\rm m}=n_{\rm m}/n_{\rm D}$ and $x_{\rm m}=n_{\rm m}/n_{\rm He}$ accordingly, where $n_{\rm D}$ and $n_{\rm He}$ are the total number densities of Deuterium and Helium.   
Equations (\ref{cin}) for reactions presented in the Table \ref{tab1} together with equations of cosmological recombinations for Hydrogen, Deuterium and Helium \citep{seager1999,seager2000,Novosyadlyj2017c} compose the system of equations of formation/recombination and dissociation/ionization of atoms and molecules which are solved numerally for each step of integration of the system (\ref{nu})-(\ref{Tm2}). We set the initial conditions for them at early epoch long before the cosmological recombination when all ingredients were completely ionized and Saha approximation was applicable (see for details the Appendix A in \citep{Novosyadlyj2017c}).  
The publicly available codes RecFast\footnote{http://www.astro.ubc.ca/people/scott/recfast.html} and DDRIV1\footnote{http://www.netlib.org/slatec/src/ddriv1.f} have been used in the general code bdedmhalo.f, which was designed for integration of the system of equations (\ref{nu})-(\ref{Tm2}) with initial conditions (\ref{r_ini})-(\ref{m_ini}) jointly with the system of equations (\ref{cin}). In the computations we assumed $f_{\rm He}=0.082$ and $f_{\rm D}=2.61\cdot10^{-5}$, that follows from the standard model of cosmological nucleosynthesis and agree with Planch2015 and other observational data. Reaction rates are taken from \citet{Galli1998}, except for the rates of recombination and photoionization of Hydrogen and Helium for $z>200$, which are taken from \citet{seager1999}.

\begin{figure*}[ht!]
\includegraphics[width=0.49\textwidth]{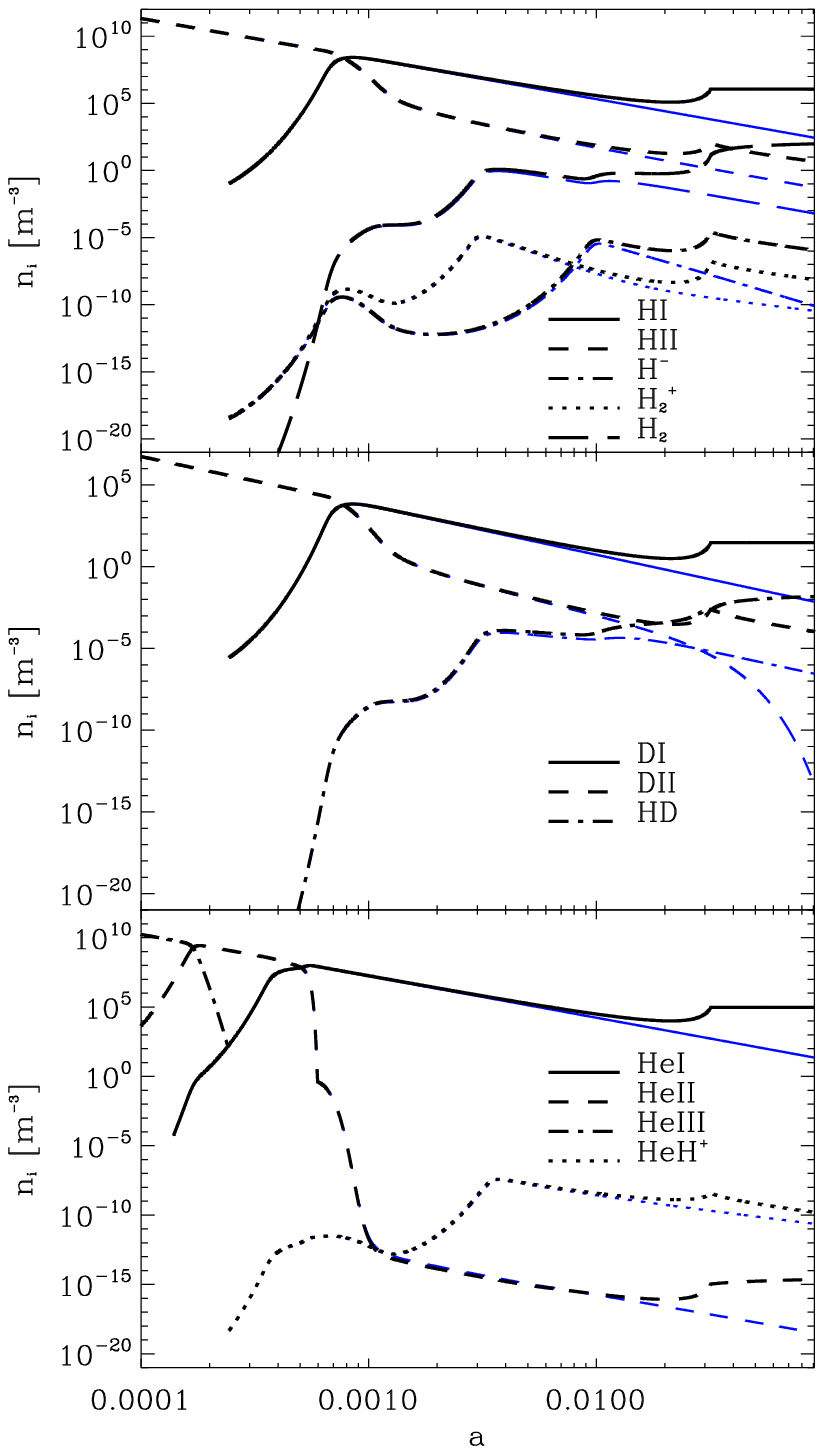}
\includegraphics[width=0.49\textwidth]{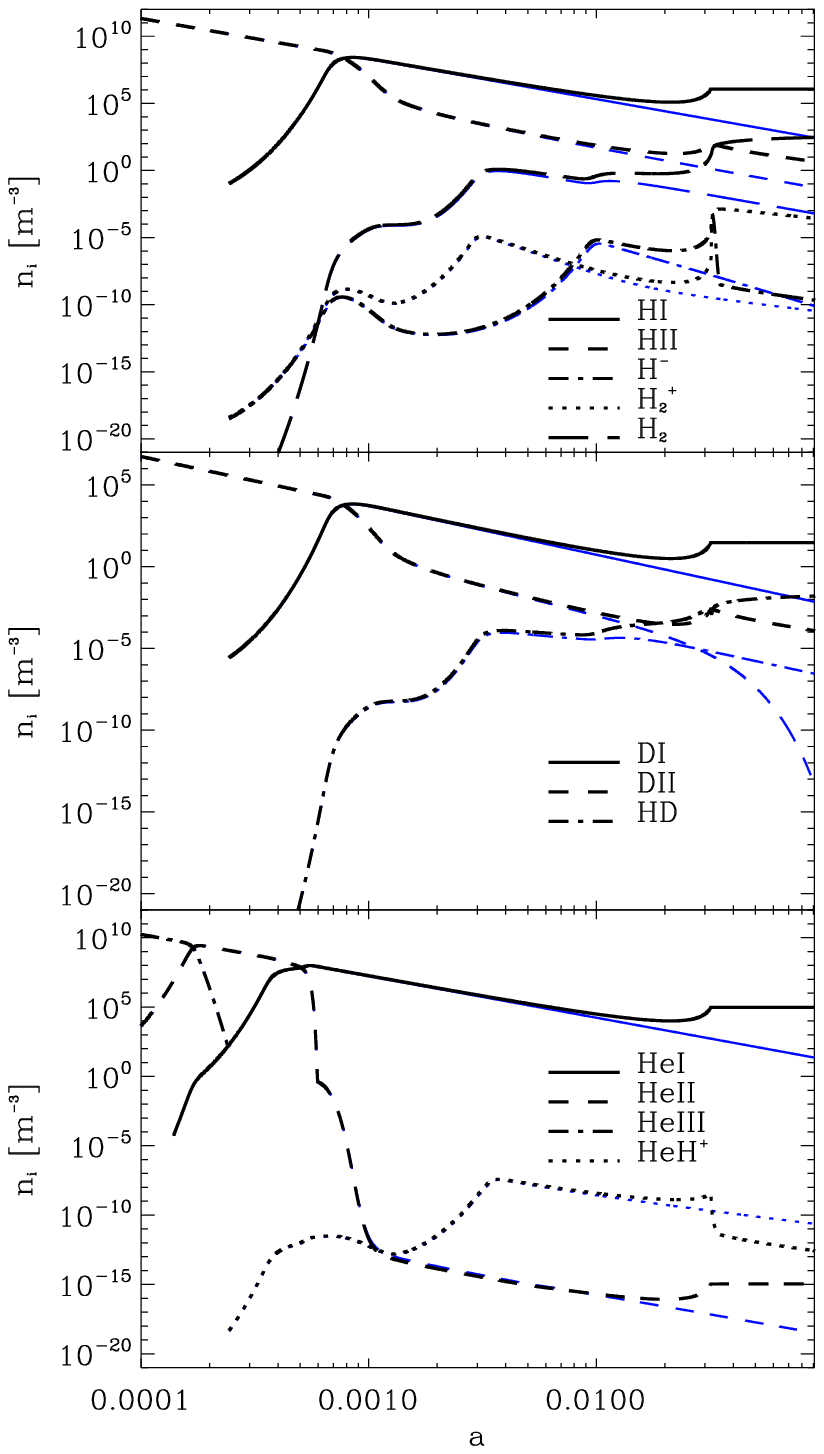} 
\caption{Evolution of the number density of atoms, ions, neutral and ionized molecules during Dark Ages on the cosmological background $n_{\rm i}^{\rm (c.b.)}$ (blue lines) and in the central part of spherical halo $n_{\rm i}^{\rm halo}$ with $M=7\cdot10^8\,\mathrm{M_{\odot}}$, which have been virialized at $z\sim30$ (black lines). In the left panel the gas in virialized halo has adiabatic temperature ($T_{\rm b}^{\rm (ad)}=400$ K) and in the right panel in the virialized halo gas has virial temperature ($T_{\rm b}^{\rm (vir)}=253800$) K.}
 \label{hdhe}
\end{figure*} 
 
The results of join integration of these systems of equations are shown in Fig. \ref{hdhe} for the halo with $M=7\cdot10^8\,\mathrm{M_{\odot}}$, initial amplitude $C=3\cdot10^{-4}$ and two cases of halo temperature: in the virialized halo the gas has adiabatic temperature $T_{\rm b}^{\rm (ad)}=400$ K (left panel) and  virial temperature $T_{\rm b}^{\rm (vir)}=253800$ K (right panel). Before virialization in both cases the temperature histories are the same. For comparison the same values for the cosmological background are shown by thin blue lines. One can see that the differences in the concentrations of molecules in the halo and in the cosmological background become noticeable in Fig. \ref{hdhe} starting with $z = 100$ and reaching several orders for some molecules at $z = 10$. 
We present also results in the numbers in the Table \ref{h2f}, where the number densities of atoms, ions and molecules on the cosmological background $n_{\rm i}^{\rm (c.b.)}$ and the ratio $n_{\rm i}^{\rm halo}/n_{\rm i}^{\rm (c.b.)}$ at different stages of halo formation: quasilinear ($z=100$), turn around ($z_{ta}\approx46$), beginning of virialization ($z_{\rm v}\approx30$) and the end of the Dark Ages ($z=10$). They show that the number density of some molecules ($\rm{H_2}$, $\rm{HD}$) and molecular ions ($\rm{H_2^+}$, $\rm{H^-}$) in halos is higher than corresponding values on the cosmological background caused by increased density and rates of reactions at quasilinear and nonlinear evolution of density and velocity perturbations. At the same time, formation of some ingredients ($\rm{HII}$, $\rm{HeII}$, $\rm{HeH^+}$) are depressed: their number densities at the moment of virialization are lower than the density contrast $\Delta_{\rm v}$. For neutral atoms HI, DI, HeI the ratio of number densities $n_{\rm i}^{\rm halo}/n_{\rm i}^{\rm (c.b.)}$ increases by $\left((z_{\rm v}+1)/11\right)^3\approx$23 times during interval from $z_{\rm v}$ to $z=10$, since fractions $x_{\rm i}$ are unchanged after virialization. For molecules, however, this ratio is other: for H$_2$ and HD, which are important coolers \citep{Puy1997}, it is $\sim140$. For some molecules this ratio depend also on heating/cooling of baryonic gas in the virialization process. In our model when gas has virial temperature the number density of negative ions H$^-$ is strongly depressed in the halo while molecular ions HeH$^+$ practically disappear. These ions have the lowest binding energies (presented in the parentheses near notation of molecules in the table) for molecules studied here.

\begin{table*}[ht]
\caption{Number densities of atoms, ions and molecules on the cosmological background $n_{\rm i}^{\rm (c.b.)}$ and ratio $n_{\rm i}^{\rm halo}/n_{\rm i}^{\rm (c.b.)}$ 
at different stages of halo formation.}
\label{h2f}
\centering
\begin{tabular}{l|l|l|l|l}
\hline
Species (binding energy) &\quad\quad $z=100$ &\quad\quad  $z=46.06$ &\quad\quad  $z=30.35$ & \hskip2.5cm $z=10$ \\
        &$n_{\rm i}^{\rm (c.b.)}\quad\quad\frac{n_{\rm i}^{\rm halo}}{n_{\rm i}^{\rm (c.b.)}}$&$n_{\rm i}^{\rm (c.b.)}\quad\quad\frac{n_{\rm i}^{\rm halo}}{n_{\rm i}^{\rm (c.b.)}}$ &$n_{\rm i}^{\rm (c.b.)}\quad\quad\frac{n_{\rm i}^{\rm halo}}{n_{\rm i}^{\rm (c.b.)}}$ &$n_{\rm i}^{\rm (c.b.)}\quad\quad\frac{n_{\rm i}^{\rm halo}}{n_{\rm i}^{\rm (c.b.)}}(T_{\rm b}^{\rm (ad)})\quad\frac{n_{\rm i}^{\rm halo}}{n_{\rm i}^{\rm (c.b.)}}(T_{\rm b}^{\rm (vir)})$ \\
\hline 
${\rm HI}$ (13.60 eV)&$2.1\cdot10^5$\hskip0.8cm1.8&$2.2\cdot10^4$\hskip0.8cm5.6&$6.3\cdot10^3$\quad\quad178&$2.8\cdot10^2$\quad\quad$4.1\cdot10^3$\quad\quad\quad$4.1\cdot10^3$\\
${\rm HII}$ &51.4\hskip1.3cm1.5&4.6\hskip1.5cm 4.0&1.3\hskip1.5cm 81&$5.4\cdot10^{-2}$\hskip0.8cm 86 \hskip1.7cm 87 \\
${\rm H^-}$ (0.754 eV) &$3.1\cdot10^{-6}$\hskip0.55cm 2.1&$1.2\cdot10^{-7}$\hskip0.6cm8.7&$1.6\cdot10^{-8}$\quad1570&$8.0\cdot10^{-11}$\quad$1.4\cdot10^4$\hskip1.3cm2.8   \\
${\rm H_2}$ (4.47 eV)&0.13\hskip1.3cm2.9&$4.6\cdot10^{-2}$\hskip0.5cm13.0&$1.4\cdot10^{-2}$\quad1010&$6.2\cdot10^{-4}$\quad\,\,$1.5\cdot10^5$\hskip1cm$4.7\cdot10^5$     \\
${\rm H_2^+}$ (2.77 eV)&$2.2\cdot10^{-8}$\hskip0.5cm1.8&$9.5\cdot10^{-10}$\hskip0.5cm4.8&$3.3\cdot10^{-10}$\quad510&$3.5\cdot10^{-11}$\quad\quad212\hskip1.3cm$7.8\cdot10^6$   \\
\hline
${\rm DI}$ (15.47 eV)&5.6\hskip1.4cm1.8&0.56\hskip1.4cm 5.6&0.17\hskip1.2cm 178&$7.2\cdot10^{-3}$\quad\,\,$4.1\cdot10^3$\hskip1cm$4.1\cdot10^3$  \\
${\rm DII}$ &$1.0\cdot10^{-3}$\hskip0.5cm1.5&$4.3\cdot10^{-5}$\hskip0.65cm7.0&$3.7\cdot10^{-6}$\quad\,\,662&$4.7\cdot10^{-14}$\quad$2.3\cdot10^9$\hskip1cm$2.6\cdot10^9$   \\
${\rm HD}$ (4.51 eV)&$3.6\cdot10^{-5}$\hskip0.5cm2.4&$2.0\cdot10^{-5}$\hskip0.5cm21.3&$6.5\cdot10^{-6}$\quad\,\,386&$2.8\cdot10^{-7}$\quad\,\,$5.3\cdot10^4$\hskip1cm $5.6\cdot10^4$     \\
\hline
${\rm HeI}$ (24.59 eV)&$1.8\cdot10^4$\hskip0.7cm1.8&$1.8\cdot10^3$\hskip0.8cm5.6&$5.3\cdot10^2$\quad\quad178&22.7\hskip1.2cm$4.1\cdot10^3$\hskip1cm$4.1\cdot10^3$   \\
${\rm HeII}$ (54.44 eV)&$1.8\cdot10^{-16}$\hskip0.4cm1.2 &$2.1\cdot10^{-17}$\hskip0.5cm4.0&$6.8\cdot10^{-18}$\quad\,159&$3.3\cdot10^{-19}$\quad\,$7.0\cdot10^3$\hskip1cm$3.3\cdot10^3$   \\
${\rm HeH^+}$ (1.85 eV)&$2.8\cdot10^{-9}$\hskip0.5cm1.4&$4.9\cdot10^{-10}$\hskip0.5cm2.7&$2.0\cdot10^{-10}$\quad\,\,\,18&$2.3\cdot10^{-11}$\hskip0.8cm 7\hskip1.5cm$1.2\cdot10^{-2}$   \\
\hline
\end{tabular}
\end{table*}
  
\section{Discussions and conclusions} 

We have analyzed the formation of spherical halos with $M\sim10^8-10^9\,\mathrm{M_{\odot}}$ which are virialized in the Dark Ages at $10\le z\le30$ in the four-component medium: dark matter, dark energy, baryons and radiation. It is shown that dark matter halos can virialize at $z\approx30$ if they are forming from high density peaks in the Gaussian field of initial density perturbations with $\delta_{\rm m}^{\rm init}\sim5-6\sigma_{\rm m}$, where $\sigma_{\rm m}$ is rms density fluctuations computed for $\Lambda$CDM model with Planck2015 parameters. The dark matter halos which are forming from peaks with $\delta_{\rm m}^{\rm init}\sim2-3\sigma_{\rm m}$ are virialized at $z\approx10$, their number density (in units Mpc$^{-3}$) are close to the number density of bright galaxies estimated on the base of galaxy redshift catalogues. The radiation component is important at the linear stage of evolution of precursors at the radiation-dominated epoch and at decoupling time. The Silk damping effect depresses the amplitude of perturbations in the baryon-photon plasma before and during cosmological recombination (Figs. \ref{vir}-\ref{vir_st}).  

The density and velocity perturbations in the dark energy component oscillate after entering the particle horizon when its effective sound speed is equal to the speed of light as it is in the case when the classical scalar field is dark energy. In this case, the dark energy perturbations are not important for dynamic of halo formation after its entering into the particle horizon (Fig. \ref{vir}). In the case when the effective sound speed is small, comparable with the effective sound speed in baryonic matter at the end of the Dark Ages for example, then such dark energy can infall into forming halo of dark matter, reaching the state of hydrostatic equilibrium (Fig. \ref{vir_st}). Such dark energy can be important also at the late stage of halo formation.  

The evolution of the baryonic component is most important since it can be observable. At the end of cosmological recombination the baryonic matter on the interested here scales have been smoothed, but immediately after recombination it starts to free fall into the potential wells of dark matter halo seeds, thus the amplitudes of density and velocity perturbations in both components become practically the same at $z\approx200$ (Fig. \ref{vir}). Further evolution of the dark matter and baryonic matter in halo is the same up to the beginning of virialization when the heating of gas makes its behavior again different. The gas in halos and on the cosmological background had different dynamical and thermal history, so, the molecular fractions can be different too. To estimate such difference we have analyzed the kinetics of formation and dissociation of molecules and molecular ions in all stages of the dark matter halo formation: the linear and quasi-linear stages when dark matter overdensity expanded and non-linear stages when it turnarounds, collapses and virializes. We have simplified the description of the last phase by stopping infall at the virial density $\rho_{\rm m}^{\rm (vir)}=\Delta_{\rm v}\rho_{\rm m}(a_{\rm v})$ (see details in \citet{Novosyadlyj2016}). The temperature of gas was equal to the temperature of radiation up to $z\approx800$, later it has been defined by adiabatic expansion before the turn around and adiabatic compression after it, that is shown in the left panel of Fig. \ref{rho-t-sj} by dark solid lines for halos virialized at $10\le z\le30$. Thin solid red lines show the virial temperatures (\ref{Tvir}) of these halos, which we set by smoothly transition from the adiabatic value to the virial one for short time corresponded $\Delta a=0.1a_{\rm v}$ immediately after $a_{\rm v}$. One can see that in any case the key parameters of virialized halo - density and temperature, - are defined by the moment of virialization $a_{\rm v}$. 

The estimated number densities of atoms, molecules and their ions in halo are essentially larger than on the cosmological background. At the moment of halo turnaround the number densities of neutral atoms HI, DI and HeI are by 5.6 times larger than corresponding values on the cosmological background. Such ratio of number densities equals to the density contrast for top-hat halo which follows from the well known Tolman model of dust-like spherical cloud. For the number density of molecules H$_2$ and HD, which are important coolers of Dark Ages protostar clouds, these ratios are 13 and 21 accordingly (Table \ref{h2f}, 3rd column). For the moment when the density contrast of collapsing halo reaches the contrast $\Delta_{\rm v}\approx178$, following from the virial theorem, the ratios $n_{\rm i}^{\rm halo}/n_{\rm i}^{\rm (c.b.)}$ for H$_2$ and HD $\approx1000$ and $\approx400$ accordingly, while for the neutral atoms HI, DI and HeI they are equal 178. This effect is explained by crucial dependence of chemical reaction chains, which lead to formation of these molecules, on the local density and temperature of Hydrogen-Deuterium slightly ionized gas. Taking into account this effect for computation of cooling/heating processes in the Dark Ages halos can help us to resolve the problem of fragmentation of primordial medium into protostar clouds with mass $\lesssim10^3\,\mathrm{M_{\odot}}$. We plan to do that in the next work. 

\acknowledgments
This work was supported by International Center of Future Science of Jilin University (P.R.China) and the project of Ministry of Education and Science of Ukraine ``Dynamics, evolution and physical effects in the formation of structure of the Universe, supernova remnants and close binary stars with relativistic components'' (state registration number 0116U001544).

\end{document}